\documentstyle[preprint,aps,epsf,]{revtex}
\begin{document}
%
%
%
\draft
%
\title{  Role of Correlation and Exchange for  Quasi-particle
Spectra of Magnetic and Diluted Magnetic Semiconductors }

%
\author{A.\ L.\ Kuzemsky\cite{KuzEmail}}
\address{Bogoliubov Laboratory of Theoretical Physics,\\
 Joint Institute for Nuclear Research, 141980 Dubna, Russia}
\date{\today}
%
\maketitle
%
%
\begin{abstract}
%
Theoretical foundation and application of the generalized
spin-fermion $(sp-d)$ exchange lattice model to magnetic and
diluted magnetic semiconductors are discussed. The capabilities of
the model to describe spin quasi-particle  spectra are
investigated.  The main emphasis is made on the dynamic behavior
of two interacting subsystems, the localized spins and spin
density of itinerant carriers. A nonperturbative many-body
approach, the Irreducible Green Functions (IGF) method, is used to
describe  the quasi-particle dynamics. Scattering states are
investigated and  three branches of magnetic excitations are
calculated  in the regime, characteristic of a magnetic
semiconductor. For a simplified version of the model ( Kondo
lattice model ) we study the spectra of quasi-particle
excitations with special attention given to diluted magnetic
semiconductors. For this, to include the effects of disorder,
modified mean fields are determined self-consistently. The role
of the Coulomb correlation and exchange is clarified by comparing
of  both the cases. 
%
\end{abstract}
%
\section{Introduction}
%
There is considerable current  interest in the properties of
magnetic\cite{di1,mau} and diluted magnetic
semiconductors\cite{di2}-\cite{di6}. This field is very active and
there are many aspects to the problem. A lot of materials were
synthesized and tested\cite{oh}-\cite{park}. The new materials
design approach to fabrication of new functional diluted magnetic
semiconductors (DMS) resulted in  producing a variety of
compounds . Diluted magnetic semiconductors are semiconducting
alloys whose lattice contains magnetic atoms as randomly
distributed substitutional impurities such as Mn-doped InAs or
GaAs ( general formula $A^{III}_{1-x}Mn_{x}B^{V} $ ). A fraction
of A sublattice which is substituted at random by Mn changes the
carrier density from the low doping (nearly insulating regime) to
the large doping (metallic regime ). The presence of the spin
degree of freedom in DMS may lead to a new semiconductor spin
electronics which will combine the advantages of the
semiconducting devices with the new features due to the
possibilities of controlling the magnetic state.\\
However, the coexistence of ferromagnetism and semiconducting
properties in these compounds require a suitable theoretical
model which would describe well both the magnetic cooperative
behavior and semiconducting properties as well as a rich field of
interplay between them.   Existence and properties of localized
and itinerant magnetism in metals, oxides and alloys and their
interplay is an interesting but not yet fully understood problem
of quantum theory of magnetism\cite{don},\cite{fed}.   Behavior
and the true nature of the electronic and spin states, and their
quasi-particle dynamics are of central importance to the
understanding of  physics of correlated systems such as magnetism
and Mott-Hubbard metal-insulator transition in metals and oxides,
magnetism and heavy fermions (HF) in rare-earths compounds, and
anomalous transport properties in perovskite manganites. This
class of systems is characterized by the complex, many-branch
spectra of elementary excitations. Moreover, the correlations
effects are essential\cite{kuz1},\cite{kuz2}.\\ Recently, there
has been considerable interest in identifying the microscopic
origin of quasi-particle states in such systems and a few  model
approaches have been proposed. Many magnetic and electronic
properties of magnetic semiconductors and related  materials may
be interpreted reasonably in terms of combined spin-fermion
models ( SFM ) which include the interacting spin and charge
subsystems\cite{don},\cite{fed},\cite{kuz1}. This approach
permits one to describe  significant and interesting physics,
e.g., the bound states and magnetic
polarons\cite{sha},\cite{kuz3}, anomalous
transport properties, etc.\\
The problem of   adequate physical description  within various
types of generalized spin-fermion model  has  intensively been
studied during the last decades, especially in the context of
magnetic and transport properties of rare-earth and transition
metals and their compounds~\cite{coq1},\cite{kuz4} and magnetic
semiconductors~\cite{kuz1}-~\cite{kuz3},\cite{tak1}-\cite{bab}.
For DMS the continuous analog of the Zener model was proposed
recently by Dietl et al.\cite{di3}. The interplay of magnetic and
semiconducting behavior was considered within a phenomenological
mean field scheme.
\\ The majority of theoretical papers on DMS
studied its properties mainly within the mean field approximation
and continuous media terms. In such a picture the disorder
effects, which play an essential role\cite{tak6}-\cite{tak8}, can
be taken into account roughly only. However, many experimental
investigations call for a better understanding of the relevant
physics and     nature of solutions (especially magnetic) in
terms of  the lattice spin-fermion model\cite{kuz4},\cite{saw}.
\\ In the previous papers, we set up the formalism of the method
of  Irreducible Green Functions (IGF)~\cite{kuz2}. This IGF
method allows one to describe the quasi-particle spectra with
damping for a many-particle system on a lattice with complex
spectra and a strong correlation in a very general and natural
way. This scheme differs from the traditional method of
decoupling of an infinite chain of  equations~\cite{tyab} and
permits construction of the relevant dynamical solutions in a
self-consistent way at the level of the Dyson equation without
decoupling the chain of  equations of motion for the GFs.\\ In
this paper, we apply the IGF formalism to consider the
quasi-particle spectra for the lattice spin-fermion model
consisting of two interacting subsystems. It is the purpose of
this paper to explore more fully the notion of Generalized Mean
Fields (GMF)~\cite{kuz2} which may arise in the system of
interacting localized spins ( including effects of disorder ) and
lattice fermions to justify and understand the "nature" of the
relevant mean fields. Theoretical foundation and application of
the generalized spin-fermion ( sp-d ) exchange model to magnetic
and diluted magnetic semiconductors are discussed in some detail.
The capabilities of the model to describe quasi-particle spectra
are investigated.  The purpose   is to investigate the
quasi-particle spectra and GMF of the  magnetic semiconductors
consisting of two interacting charge and spin subsystems within
the lattice spin-fermion model in a unified and coherent fashion
to analyze the role and influence of the Coulomb correlation and
exchange.
%
\section{Spin-Fermion Model}
%
 The concept of the $sp-d$ ( or $d-f$ )
model plays an important role in the quantum theory of
magnetism\cite{don},\cite{fed},\cite{kuz4},~\cite{kri},\cite{lar},\cite{coq1}.
In this section, we consider the generalized $sp-d$ model  which
describes the localized $3d(4f)$-spins interacting with
$s(p)$-like conduction (itinerant) electrons ( or holes ) and
takes into consideration the electron-electron interaction.
 \\ The total Hamiltonian of the model
is given by \begin{equation} \label{eq1}
 H = H_{s} + H_{s-d} + H_{d}
\end{equation}
The Hamiltonian of band electrons (or holes) is given by
\begin{equation}
\label{eq2} H_{s} = \sum_{ij} \sum_{\sigma}
t_{ij}a^{\dagger}_{i\sigma}a_{j\sigma} + \frac{1}{2}
\sum_{i\sigma} Un_{i\sigma}n_{i-\sigma}
\end{equation}
This is the Hubbard model. We adopt the notation
$$a_{i\sigma} =
N^{-1/2}\sum_{\vec k}  a_{k\sigma} \exp (i{\vec k} {\vec R_{i}})
\quad a^{\dagger}_{i\sigma} = N^{-1/2}\sum_{\vec k}
a^{\dagger}_{k\sigma} \exp (-i{\vec k} {\vec R_{i}})$$
In the case of a pure semiconductor, at low temperatures the
conduction electron band is empty and the Coulomb term $U$ is
therefore not so important. A partial occupation of the band
leads to an increase in the role of the Coulomb correlation. It
is clear that  we treat conduction electrons as s-electrons in the
Wannier representation. In doped DMS the
carrier system is the valence band p-holes.\\
 The band energy of Bloch electrons
$\epsilon(\vec k)$ is defined as follows: $$t_{ij} =
N^{-1}\sum_{\vec k}\epsilon(\vec k) \exp[i{\vec k}({\vec R_{i}}
-{\vec R_{j}})],$$ where  $N$ is the number of  lattice sites.
For the tight-binding electrons in a cubic lattice we use the
standard expression for the dispersion
\begin{equation}\label{eq3}
\epsilon(\vec k) = 2\sum_{\alpha}t( \vec a_{\alpha})\cos(\vec k
\vec a_{\alpha}) \end{equation}, where $\vec a_{\alpha}$ denotes
the lattice vectors in a simple lattice with the inversion centre.\\
The term $H_{s-d}$ describes the interaction of the total
3d(4f)-spin  with the spin density of the itinerant
carriers\cite{lar}
\begin{equation} \label{eq4}
H_{s-d} = -2\sum_{i}I{\vec \sigma_{i}}{\vec S_{i}} = - I
N^{-1/2}\sum_{kq}\sum_{\sigma}[S^{-\sigma}_{-q}a^{\dagger}_{k\sigma}
a_{k+q-\sigma} +
z_{\sigma}S^{z}_{-q}a^{\dagger}_{k\sigma}a_{k+q\sigma}]
\end{equation}
where sign factor $z_{\sigma}$ is given by $$z_{\sigma} = (+ or
-)\quad for \quad \sigma =  (\uparrow  or  \downarrow)$$ and
$$S^{-\sigma}_{-q} = \cases {S^{-}_{-q} &if $\sigma = +$ \cr
S^{+}_{-q} &if $\sigma = -$ \cr}$$.\\
In DMS the local exchange coupling resulted from the $p-d$
hybridization between the Mn $d$ levels and the $p$ valence band
$I \sim V_{p-d}^2$ . For the subsystem of
  localized spins we have
\begin{equation}
\label{eq5}
 H_{d} = -\frac{1}{2} \sum_{ij} J_{ij} \vec S_{i}\vec S_{j} =
-\frac{1}{2} \sum_{q} J_{q} \vec S_{q}\vec S_{-q}
\end{equation}
Here we use the notation
$$S_{i}^{\alpha} =
N^{-1/2}\sum_{\vec k}  S_{k}^{\alpha}  \exp (i{\vec k} {\vec
R_{i}}) \quad  S_{k}^{\alpha} = N^{-1/2}\sum_{\vec k}
 S_{i}^{\alpha} \exp (-i{\vec k} {\vec R_{i}})$$
$$ [ S_{k}^{\pm},S_{q}^{z}] = \mp S_{k+q}^{\pm} \quad
[ S_{k}^{+},S_{q}^{-}] =  \frac{2}{N^{1/2}} S_{k+q}^{z}$$
$$J_{ij} =
N^{-1}\sum_{\vec k}J_{\vec k}  \exp[i{\vec k}({\vec R_{i}} -{\vec
R_{j}})],$$
 This term describes the direct exchange
interaction between the localized 3d (4f) magnetic moments at the
lattice sites i and j. In the DMS system this interaction is
rather small. The ferromagnetic interaction between the local Mn
moments is mediated by the real itinerant carriers in the valence
band of the host semiconductor material. The carrier polarization
produces the RKKY exchange interaction of Mn local moments
\begin{equation}
\label{eq5a}
 H_{RKKY} =  - \sum_{i\neq j} K_{ij}  \vec S_{i}\vec S_{j}
\end{equation}
We emphasize that $ K_{ij}\sim |I^2| \sim V_{p-d}^4$. To explain
this, let us remind that the microscopic model\cite{lar}, which
contains the basic physics, is the Anderson-Kondo model
\begin{eqnarray}
\label{eq5b} H  = \sum_{ij} \sum_{\sigma}
t_{ij}a^{\dagger}_{i\sigma}a_{j\sigma} - V\sum_{ij}
\sum_{\sigma}(a^{+}_{i\sigma}d_{j\sigma} + h.c.)
\nonumber \\
-E_{d}\sum_{i} \sum_{\sigma} n^{d}_{i\sigma} + \frac{1}{2}
\sum_{i\sigma} Un^{d}_{i\sigma}n^{d}_{i-\sigma}
\end{eqnarray}
For the symmetric case $U = 2E_{d}$ and for $U \gg V$
Eq.(\ref{eq5b}) can be mapped onto  the Kondo lattice model ( KLM
)
\begin{equation}
\label{eq5c}
 H  = \sum_{ij} \sum_{\sigma}
t_{ij}a^{\dagger}_{i\sigma}a_{j\sigma} - \sum_{i}2I{\vec
\sigma_{i}}{\vec S_{i}}
\end{equation}
Here $I \sim \frac {4V^{2}}{E_{d}}$. The KLM may be viewed as the
low-energy sector of the initial model Eq.( \ref{eq5b} ).

\section{ Outline of the IGF Method} In this section, we  discuss
the main ideas of the IGF approach that allows one to describe
completely  quasi-particle spectra with damping in a very natural way.\\
We  reformulated  the two-time GF method~\cite{kuz2} to the form
which is especially adjusted  to correlated fermion systems on a
lattice and systems with complex spectra~\cite{kuz1}-\cite{kuz4}.
A very important concept of the whole method is the {\it
Generalized Mean Fields} (GMFs), as it was formulated
in~\cite{kuz2}. These GMFs have a complicated structure for the
strongly correlated case and complex spectra, and are not reduced
to the functional of  mean densities of the electrons or spins
when
one calculates excitation spectra at finite temperatures. \\
To clarify the foregoing, let us consider a retarded GF of the
form~\cite{tyab}
\begin{equation}
\label{eq6} G^{r} = <<A(t), A^{\dagger}(t')>> = -i\theta(t -
t')<[A(t) A^{\dagger}(t')]_{\eta}>, \eta = \pm 1
\end{equation}
As an introduction to the concept of IGFs, let us describe the
main ideas of this approach in a symbolic and simplified form. To
calculate the retarded GF $G(t - t')$,  let us write down the
equation of motion for it
\begin{equation}
\label{eq7} \omega G(\omega) = <[A, A^{\dagger}]_{\eta}> + <<[A,
H]_{-}\mid A^{\dagger}>>_{\omega}
\end{equation} The essence of the method
is as follows~\cite{kuz2}: \\ It is based on the notion of the
{\it "IRREDUCIBLE"} parts of GFs (or the irreducible parts of the
operators, $A$ and $A^{\dagger}$, out of which the GF is
constructed) in terms of which it is possible, without recourse
to a truncation of the hierarchy of equations for the GFs, to
write down the exact Dyson equation and to obtain an exact
analytic representation for the self-energy operator. By
definition, we introduce the irreducible part {\bf (ir)} of the GF
\begin{equation}
\label{eq8} ^{(ir)}<<[A, H]_{-}\vert A^{\dagger}>> = <<[A, H]_{-}
- zA\vert A^{\dagger}>>
\end{equation}
The unknown constant z is defined by the condition (or constraint)
\begin{equation}
\label{eq9} <[[A, H]^{(ir)}_{-}, A^{\dagger}]_{\eta}> = 0
\end{equation}
which is an analogue of the orthogonality condition in the Mori
formalism. From the condition (\ref{eq9}) one can find
\begin{equation}
\label{eq10} z = \frac{<[[A, H]_{-}, A^{\dagger}]_{\eta}>}{<[A,
A^{\dagger}]_{\eta}>} =
 \frac{M_{1}}{M_{0}}
\end{equation}
Here $M_{0}$ and $M_{1}$ are the zeroth and first order moments
of the spectral density. Therefore, the irreducible GFs  are
defined so that they cannot be reduced to the lower-order ones by
any kind of decoupling. It is worth  noting that the term {\it
"irreducible"} in a group theory means a representation of a
symmetry operation that cannot be expressed in terms of lower
dimensional representations. Irreducible (or connected )
correlation functions are known in statistical mechanics. In the
diagrammatic approach, the irreducible vertices are defined as
graphs that do not contain inner parts connected by the
$G^{0}$-line. With the aid of the definition (\ref{eq8})   these
concepts are expressed in terms  of retarded and advanced GFs.
The procedure extracts all relevant (for the problem under
consideration) mean-field contributions and puts them into the
generalized mean-field GF which  is defined here as
\begin{equation}
\label{eq11} G^{0}(\omega) = \frac{<[A,
A^{\dagger}]_{\eta}>}{(\omega - z)}
\end{equation}
To calculate the IGF $\quad  ^{(ir)}<<[A, H]_{-}(t),
A^{\dagger}(t')>>$ in (\ref{eq7}), we have to write the equation
of motion for it after differentiation with respect to the second
time variable $t'$. The condition of orthogonality (\ref{eq9})
removes the inhomogeneous term from this equation and is a very
crucial point of the whole approach. If one introduces the
irreducible part for the right-hand side operator as discussed
above for the ``left" operator, the equation of motion
(\ref{eq7}) can be exactly rewritten in the following form
\begin{equation} \label{eq12} G = G^{0} + G^{0}PG^{0}
\end{equation} The scattering operator $P$ is given by
\begin{equation}
\label{eq13} P = (M_{0})^{-1}(\quad ^{(ir)}<<[A,
H]_{-}\vert[A^{\dagger}, H]_{-}>>^{(ir)}) (M_{0})^{-1}
\end{equation}
The structure of  equation ( \ref{eq13}) enables us to determine
the self-energy operator $M$, by  analogy with the diagram
technique
\begin{equation} \label{eq14} P = M + MG^{0}P
\end{equation}
From the definition (\ref{eq14}) it follows that  the self-energy
operator $M$ is defined as a proper (in the diagrammatic language,
``connected") part of the scattering operator $M = (P)^{p}$. As a
result, we obtain the exact Dyson equation for the thermodynamic
double-time Green functions
\begin{equation} \label{eq15} G =
G^{0} + G^{0} M G
\end{equation}
The difference between $P$ and $M$ can be regarded as two
different solutions of two integral equations (\ref{eq12}) and
(\ref{eq15}). But from  the Dyson equation (\ref{eq15}) only the
full GF  is seen to be expressed as a  formal solution of the form
\begin{equation}
\label{eq16} G = [ (G^{0})^{-1} - M ]^{-1}
\end{equation}
Equation  (\ref{eq16}) can be regarded as an alternative form of
the Dyson equation (\ref{eq15}) and the {\it definition} of $M$
provided that the generalized mean-field GF $G^{0}$ is specified.
On the contrary , for the scattering operator $P$, instead of the
property $G^{0}G^{-1} + G^{0}M = 1$, one has the property
$$(G^{0})^{-1} - G^{-1} = P G^{0}G^{-1}$$  Thus, the { \it very functional
form} of the formal solution (\ref{eq16}) determines the
difference between $P$ and $M$ precisely. \\ Thus, by introducing
irreducible parts of GF (or  irreducible parts of the operators,
out of which the GF is constructed) the equation of motion
(\ref{eq7}) for the GF can exactly be  ( but using the
orthogonality constraint (\ref{eq9})) transformed into the Dyson
equation for the double-time thermal GF (\ref{eq15}). This result
is very remarkable  because  the traditional form of the GF
method does not include this point. Notice that all quantities
thus considered are  exact. Approximations can be generated not
by truncating the set of coupled equations of motions but by a
specific approximation of the functional form of the mass
operator $M$ within a self-consistent scheme  expressing $M$ in
terms of initial GF
$$ M \approx F[G]$$
Different approximations are relevant to different physical
situations.\\The projection operator technique  has essentially
the same philosophy. But with using the constraint (\ref{eq9}) in
our approach we emphasize the fundamental and central role of the
Dyson equation for the calculation of single-particle properties
of  many-body systems. The problem of reducing the whole
hierarchy of equations involving higher-order GFs by a coupled
nonlinear set of integro-differential equations connecting the
single-particle GF to the self-energy operator is rather
nontrivial. A characteristic feature of these equations is that,
besides the single-particle GF, they involve also higher-order
GF. The irreducible counterparts of the GFs, vertex functions,
serve to identify correctly the self-energy as
$$  M = G^{-1}_{0}  - G^{-1}$$
The integral form of Dyson equation (\ref{eq15}) gives  $M$ the
physical meaning of a nonlocal and energy-dependent effective
single-particle potential. This meaning can be verified for the
exact self-energy through the diagrammatic expansion for the
causal GF.\\
It is important to note that for the retarded and advanced GFs,
the notion of the proper part $M = (P)^{p}$ is symbolic in
nature~\cite{kuz2}. In a certain sense, it is possible to say that
it is defined here by analogy with the irreducible many-particle
$T$-matrix. Furthermore, by analogy with the diagrammatic
technique, we can also introduce the proper part defined as a
solution to the integral equation (\ref{eq14}). These analogues
allow us to understand better the formal structure of the Dyson
equation for the double-time thermal GF but only in a symbolic
form . However, because of the identical form of the equations
for  GFs for all three types ( advanced, retarded, and causal ),
we can convert in each stage of calculations to causal GF and,
thereby, confirm the substantiated nature of definition
(\ref{eq14})! We therefore should speak of an analogy of the Dyson
equation. Hereafter, we  drop this stipulating, since it does not
cause any misunderstanding. In a sense, the IGF method is a
variant of the Gram-Schmidt orthogonalization procedure.\\It
should be emphasized that the scheme presented above gives just a
general idea of the IGF method. A more exact explanation why one
should not introduce the approximation already in $P$, instead of
having to work out $M$, is given below when working out the
application
of the method to  specific problems.\\
The general philosophy of the IGF method is in the separation and
identification of elastic scattering effects and inelastic ones.
This latter point is quite often underestimated, and both effects
are mixed. However, as far as the right definition of
quasi-particle damping is concerned, the separation of elastic
and inelastic scattering processes is believed to be crucially
important for  many-body systems with complicated spectra and
strong interaction.   \\ From a technical point of view, the
elastic GMF renormalizations can exhibit  quite a nontrivial
structure. To obtain this structure correctly, one should
construct the full GF from the complete algebra of  relevant
operators and develop a special projection procedure for
higher-order GFs, in accordance with a given algebra. Then a
natural question arises how to select the relevant set of
operators $\{ A_{1}, A_{2}, ... A_{n} \}$   describing the
"relevant degrees of freedom". The above consideration suggests
an intuitive and heuristic way to the suitable procedure as
arising from an infinite chain of equations of motion
(\ref{eq7}). Let us consider the column
$$ \pmatrix{ A_{1}\cr  A_{2}\cr \vdots \cr  A_{n}\cr}$$
where
$$ A_{1} = A,\quad A_{2} = [A,H],\quad A_{3} = [[A,H],H], \ldots
A_{n} = [[... [A, \underbrace{H]...H}_{n}]$$ Then the most
general possible Green function can be expressed as a matrix
$$ \hat G = <<\pmatrix{
A_{1}\cr  A_{2}\cr \vdots \cr A_{n}\cr} \vert \pmatrix{
A^{\dagger}_{1}& A^{\dagger} _{2}& \ldots &  A^{\dagger}
_{n}\cr}>>$$ This generalized Green function describes the one-,
two- and $n$-particle dynamics. The equation of motion for it
includes, as a particular case, the Dyson equation for
single-particle Green function, and the Bethe-Salpeter equation
which is the equation of motion for the two-particle Green
function and which is an analogue of the Dyson equation, etc . The
corresponding reduced equations should be extracted from the
equation of motion for the generalized GF with the aid of the
special techniques such as the projection method and similar
techniques. This must be a final goal towards a real
understanding of the true many-body dynamics. At this point, it
is worthwhile to underline that the above discussion is a
heuristic scheme only but not a straightforward recipe. The
specific method of introducing  the IGFs depends on the form of
operators $A_{n}$, the type of the Hamiltonian, and conditions of
the problem.\\
Here a sketchy form of the IGF method is presented. The aim is to
introduce the general scheme and to lay the groundwork for
generalizations.   We  demonstrated  in~\cite{kuz2} that the IGF
method is a powerful tool for describing the quasi-particle
excitation spectra, allowing a deeper understanding of elastic
and inelastic quasi-particle scattering effects and the
corresponding aspects of damping and finite lifetimes. In the
present context, it provides an efficient tool for  analysis of
the mean fields and generalized mean fields of the complicated
many-body models.
%
%
\section{Quasi-particle Dynamics of the $(sp-d)$ Model}
To describe self-consistently the spin dynamics of the extended
$sp-d$ model, one should take into account the full algebra of
relevant operators of the suitable "spin modes"  which are
appropriate when the goal is to describe self-consistently the
quasi-particle spectra of two interacting subsystem.\\
We have two kinds of spin variables
$$ S^{+}_{k}, \quad S^{-}_{-k} = ( S^{+}_{k} )^{\dagger}$$
$$ \sigma^{+}_{k} = \sum_{q} a^{+}_{q\uparrow}a_{k+q\downarrow},
\quad \sigma^{-}_{-k} = ( \sigma^{+}_{k} )^{\dagger} = \sum_{q}
a^{+}_{k+q\downarrow}a_{q\uparrow}$$ Let us consider the
equations of motion
\begin{equation}
\label{eq17}  [ S^{+}_{k}, H_{s-d}]_{-} = - I N^{-1}\sum_{pq} [
2S^{z}_{k-q}a^{\dagger}_{p\uparrow}a_{p+q\downarrow} - S^{+
}_{k-q} ( a^{\dagger}_{p\uparrow} a_{p+q\uparrow} -
a^{\dagger}_{p\downarrow} a_{p+q \downarrow})]
\end{equation}
\begin{equation}
\label{eq18}  [ S^{-}_{-k}, H_{s-d}]_{-} = - I N^{-1}\sum_{pq} [
2S^{z}_{k-q}a^{\dagger}_{p\downarrow}a_{p+q\uparrow} - S^{-}_{k-q}
( a^{\dagger}_{p\uparrow} a_{p+q-\uparrow} -
a^{\dagger}_{p\downarrow} a_{p+q \downarrow})]
\end{equation}
\begin{equation}
\label{eq19}  [ S^{z}_{k}, H_{s-d}]_{-} = - I N^{-1}\sum_{pq} (
S^{+}_{k-q}a^{\dagger}_{p\downarrow}a_{p+q\uparrow} - S^{-}_{k-q}
a^{\dagger}_{p\uparrow} a_{p+q\downarrow}  )
\end{equation}
\begin{equation}
\label{eq20}  [ S^{+}_{k}, H_{d}]_{-} =   N^{-1/2}\sum_{q} J_{q}(
S^{z}_{q}S^{+ }_{k-q}  - S^{z}_{k-q}S^{+}_{q})
\end{equation}
\begin{equation}
\label{eq21}  [ S^{-}_{-k}, H_{d}]_{-} =   N^{-1/2}\sum_{q} J_{q}(
S^{z}_{-(k+q)}S^{-}_{q}  - S^{z}_{q}S^{-}_{-(k+q)})
\end{equation}
\begin{eqnarray}
\label{eq22}  [ a^{\dagger}_{q\uparrow}a_{q+k \downarrow},
H_{s}]_{-} = ( \epsilon(q+k) -
\epsilon(q))a^{\dagger}_{q\uparrow}a_{q+k \downarrow} + \nonumber \\
UN^{-1}\sum_{pp'} ( a^{\dagger}_{q\uparrow}a^{\dagger}_{p+p'
\uparrow}a _{p\uparrow}a_{q+p'+k \downarrow} -
a^{\dagger}_{q+p'\uparrow}a^{\dagger}_{p-p' \downarrow}a
_{p\downarrow}a_{q+k \downarrow} )
\end{eqnarray}
\begin{eqnarray}
\label{eq23}  [ a^{\dagger}_{q\uparrow}a_{q+k \downarrow},
H_{s-d}]_{-} = IN^{-1/2}\sum_{pp'} [S^{+}_{-p'}(
a^{\dagger}_{q\uparrow}a_{p+p' \uparrow}\delta_{p,q+k} -
a^{\dagger} _{p\downarrow}a_{q+k
\downarrow}\delta_{q,p+p'}) \nonumber \\
 - S^{z}_{-p'} ( a^{\dagger}_{q\uparrow}a_{p+p'
\downarrow}\delta_{p,q+k} + a^{\dagger} _{p\uparrow}a_{q+k
\downarrow}\delta_{q,p+p'})]
\end{eqnarray}
From   Eq.(\ref{eq17}) - Eq.(\ref{eq23}) it follows that the
localized and itinerant spin variables are coupled. Suitable
algebra of relevant operators should be described by the 'spinor'
${\vec S_{i}\choose \vec \sigma_{i}}$ ("relevant degrees of
freedom"), according to the IGF strategy. \\
The model Hamiltonian $ H = H_{s} + H_{s-d} + H_{d}$ was used
in\cite{bar},\cite{kis} for   calculations of the spin-wave
spectra and was called the modified Zener model. In this model,
as applied to transition metals, the itinerant electrons are
described by a Hubbard Hamiltonian and the itinerant electron
couples   the localized spin (Hund's rule coupling) by a term
$H_{s-d}$. Because of the inequivalent spin systems, localized
and itinerant, a consequence of the model is the existence of
acoustic and optic branches of the quasi-particle spectrum of spin
excitations. In DMS the local antiferromagnetic interaction
$H_{s-d}$ produces the coupling between the carriers ( which are
holes in GaMnAs ) and the Mn magnetic moments ( s = 5/2) which
leads to ferromagnetic ordering of Mn spins in a certain range of
concentration. The Kondo physics is irrelevant in this case, but
the fully determined and consistent microscopic mechanism of the
ferromagnetic ordering is still under debates\cite{di2}. An
important question in this context is the self-consistent picture
of the quasi-particle many-body dynamics which takes into account
the complex structure of the spectra.
\subsection{Spin Dynamics of the $s-d$ Model. Scattering Regime.}
In this section, we discuss the spectrum of spin excitations in
the $sp-d$ model. We consider the double-time thermal GF of
localized spins~\cite{tyab}  which is defined as
\begin{eqnarray}\label{eq24} G^{+-}(k;t - t') =
<<S^{+}_{k}(t),S^{-}_{-k}(t')>> = -i\theta(t -
t')<[S^{+}_{k}(t),S^{-}_{-k}(t')]_{-}> = \nonumber\\ 1/2\pi
\int_{-\infty}^{+\infty} d\omega \exp(-i\omega t)
G^{+-}(k;\omega) \end{eqnarray} The next step is to write down
the equation of motion for the GF. Our attention will be focused
on spin dynamics of the model. To describe self-consistently of
the spin dynamics of the $sp-d$ model, one should take into
account the full algebra of relevant operators of the suitable
"spin modes"  which are appropriate when the goal is to describe
self-consistently the quasi-particle spectra of two interacting
subsystems. We   introduce the generalized matrix GF of the form
\begin{equation}\label{eq25}
\pmatrix{ <<S^{+}_{k}\vert S^{-}_{-k}>> & <<S^{+}_{k}\vert
\sigma^{-}_{-k}>> \cr <<\sigma^{+}_{k}\vert S^{-}_{-k}>> &
<<\sigma^{+}_{k}\vert \sigma^{-}_{-k}>> \cr} = \hat G(k;\omega)
\end{equation}
Here $$\sigma^{+}_{k} = \sum_{q}
a^{\dagger}_{k\uparrow}a_{k+q\downarrow} ;\quad \sigma^{-}_{k} =
\sum_{q} a^{\dagger}_{k\downarrow}a_{k+q\uparrow} $$
Equivalently, we can do the calculations with the matrix of the
form
\begin{equation}\label{eq26}
\pmatrix{ <<S^{+}_{k}\vert S^{-}_{-k}>> & <<S^{+}_{k}\vert
a^{\dagger}_{k+q\downarrow}a_{q\uparrow}>> \cr <<
a^{\dagger}_{q\uparrow}a_{q+k \downarrow}\vert S^{-}_{-k}>> & <<
a^{\dagger}_{q\uparrow}a_{q+k \downarrow} \vert
a^{\dagger}_{k+q\downarrow}a_{q\uparrow} >> \cr} = \hat
G'(k;\omega),
\end{equation}
but the form of Eq.(\ref{eq25}) is slightly more convenient.\\Let
us consider the equation of motion for the GF $\hat G(k;\omega)$.
By differentiation  of the GF $<<S^{+}_{k}(t) \vert B (t')>> $
with respect to the first time, $t$, we find
\begin{eqnarray}\label{eq27}
\omega<<S^{+}_{k} \vert B >>_{\omega} =
{2N^{-1/2}<S^{z}_{0}>\brace 0} + \\ \nonumber \frac {I}{N}
\sum_{pq} << S^{+}_{k-q}(a^{\dagger}_{p\uparrow}a_{p+q\uparrow} -
a^{\dagger}_{p\downarrow}a_{p+q\downarrow}) -
2S^{z}_{k-q}a^{\dagger}_{p\uparrow}a_{p+q\downarrow}\vert
B>>_{\omega} \nonumber\\  + N^{-1/2}\sum_{q} J_{q}<<(
S^{z}_{q}S^{+ }_{k-q}  - S^{z}_{k-q}S^{+}_{q})\vert B>>_{\omega}
\end{eqnarray}
where
$$ B = {S^{-}_{-k} \brace \sigma^{-}_{-k}}
$$\\
Let us introduce by definition irreducible $(ir)$ operators as
\begin{eqnarray}\label{eq28}
(S^{z}_{q})~^{ir} = S^{z}_{q} -<S^{z}_{0}>\delta_{q,0}; \quad
(a^{\dagger}_{p+q\sigma}a_{p\sigma})~^{ir} =
a^{\dagger}_{p+q\sigma}a_{p\sigma} -
<a^{\dagger}_{p\sigma}a_{p\sigma}>\delta_{q,0}\\
( (S^{z}_{q})~^{ir}S^{+ }_{k-q}  -
(S^{z}_{k-q})~^{ir}S^{+}_{q})~^{ir} =  ((S^{z}_{q})~^{ir}S^{+
}_{k-q}  - (S^{z}_{k-q})~^{ir}S^{+}_{q}) - (\phi_{q} -
\phi_{k-q})S^{+}_{k}
\end{eqnarray}
From the condition (\ref{eq9})
$$<[(  (S^{z}_{q})~^{ir}S^{+}_{k-q}  -
(S^{z}_{k-q})~^{ir}S^{+}_{q} - ( \phi_{q} - \phi_{k-q})S^{+}_{k}
), S^{-}_{-k} ]_{-}> = 0 $$ one can find
\begin{eqnarray}\label{eq29}
\phi_{q} = \frac{2K^{zz}_{q} + K^{-+}_{q}}{2<S^{z}_{0}>}\\
K^{zz}_{q} = <(S^{z}_{q})~^{ir}(S^{z}_{q})~^{ir}>; \quad
K^{-+}_{q} = <S^{-}_{-q}S^{+}_{q}>
\end{eqnarray}
 Using the definition of the irreducible parts the equation
of motion Eq.(\ref{eq27} ) can be exactly transformed to the
following form:
\begin{equation} \label{eq30} \Omega_{1}<<S^{+}_{k} \vert B
>>_{\omega} +  \Omega_{2}<<\sigma^{+}_{k} \vert
B>>_{\omega} =       {(\frac{N^{1/2}}{I})\Omega_{2}\brace 0}  +
<<A_{1} \vert B>>_{\omega}
\end{equation}
where
\begin{equation} \label{eq31} \Omega_{1} = \omega -
\frac {<S^{z}_{0}>}{N^{1/2}} ( J_{0} - J_{k} ) -  N^{-1/2}
\sum_{q}( J_{q} - J_{q-k} )\frac{2K^{zz}_{q} +
K^{-+}_{q}}{2<S^{z}_{0}>} -  I ( n_{\uparrow} - n_{\downarrow})
\end{equation}
\begin{equation} \label{eq32} \Omega_{2} =
\frac {2<S^{z}_{0}>I}{N }
\end{equation}
$$n_{\sigma} = \frac{1}{N}\sum_{q}  <a^{\dagger}_{q\sigma}a_{q\sigma}>  =
\frac{1}{N} \sum_{q} f_{q\sigma}
=\sum_{q}(\exp(\beta\epsilon(q\sigma)) + 1) $$
$$\epsilon(q \sigma) = \epsilon(q) -z_{\sigma}IN^{-1/2}<S^{z}_{0}>
+  U n_{-\sigma} $$
$$ \bar n = \sum  ( n_{\uparrow} + n_{\downarrow}); \quad 0 \le
\bar n \le 2 $$
The many-particle operator $A_{1}$ reads
\begin{eqnarray}\label{eq33}
A_{1} =
 \frac {I}{N}
\sum_{pq} [ S^{+}_{k-q}(a^{\dagger}_{p\uparrow}a_{p+q\uparrow} -
a^{\dagger}_{p\downarrow}a_{p+q\downarrow})~^{ir} -
2(S^{z}_{k-q})~^{ir}a^{\dagger}_{p\uparrow}a_{p+q\downarrow}]
  \nonumber\\  + N^{-1/2}\sum_{q}  J_{q}  (
(S^{z}_{q})~^{ir}S^{+ }_{k-q}  -
(S^{z}_{k-q})~^{ir}S^{+}_{q})~^{ir}
\end{eqnarray}
and it satisfies the conditions
$$ <[A_{1},S^{-}_{-q}]_{-}> = <[A_{1},\sigma^{-}_{-q}]_{-}> = 0$$
 To write down the equation of
motion for the Fourier transform of the GF
$<<~\sigma^{+}_{k}(t),B(t')>>$, we need the following auxiliary
equation of motion:
\begin{eqnarray}\label{eq34}
(\omega + \epsilon(p) - \epsilon(p + k) - 2IN^{-1/2}<S^{z}_{0}> -
U(n_{\uparrow} - n_{\downarrow}))
<<a^{\dagger}_{p\uparrow}a_{p+k\downarrow} \vert B >>_{\omega} +\\
\nonumber UN^{-1}(f_{p\uparrow} - f_{p+k\downarrow})
<<\sigma^{+}_{k} \vert B>>_{\omega} + IN^{-1/2}(f_{p\uparrow} -
f_{p+k\downarrow})<<S^{+}_{k} \vert
B>>_{\omega} =\\
\nonumber {0 \brace (f_{p\uparrow} - f_{p+k\downarrow})} -
IN^{-1/2} \sum_{qr}
<<S^{+}_{-r}(a^{\dagger}_{p\uparrow}a_{q+r\uparrow} \delta_{p+k,q}
- a^{\dagger}_{q\downarrow}a_{p+k\downarrow} \delta_{p,q+r})^{~ir}
\vert B>>_{\omega} - \\
\nonumber IN^{-1/2} \sum_{qr}
<<(S^{z}_{-r})^{~ir}(a^{\dagger}_{q\uparrow}a_{p+k\downarrow}
\delta_{p,q+r} + a^{\dagger}_{p\uparrow}a_{q+r\downarrow}
\delta_{p+k,q}) \vert B>>_{\omega} + \\
\nonumber UN^{-1} \sum_{qr}
<<(a^{\dagger}_{p\uparrow}a^{\dagger}_{q+r\uparrow}a_{q\uparrow}a_{p+r+k\downarrow}
-
a^{\dagger}_{p+r\uparrow}a^{\dagger}_{q-r\downarrow}a_{q\downarrow}a_{p+k\downarrow})^{~ir}
\vert B>>_{\omega}
\end{eqnarray}
Let us use  the following notation:
\begin{eqnarray}\label{eq35}
\label{eq36} A_{2} = -IN^{-1/2} \sum_{qr}
[S^{+}_{-r}(a^{\dagger}_{p\uparrow}a_{q+r\uparrow} \delta_{p+k,q}
- a^{\dagger}_{q\downarrow}a_{p+k\downarrow}
\delta_{p,q+r})^{~ir} - \nonumber\\
(S^{z}_{-r})^{~ir}(a^{\dagger}_{q\uparrow}a_{p+k\downarrow}
\delta_{p,q+r} + a^{\dagger}_{p\uparrow}a_{q+r\downarrow}
\delta_{p+k,q}) ] + \\
\nonumber UN^{-1} \sum_{qr}
(a^{\dagger}_{p\uparrow}a^{\dagger}_{q+r\uparrow}a_{q\uparrow}a_{p+r+k\downarrow}
-
a^{\dagger}_{p+r\uparrow}a^{\dagger}_{q-r\downarrow}a_{q\downarrow}a_{p+k\downarrow})^{~ir}\\
\label{eq39}
\omega_{p,k} = (\omega + \epsilon(p) - \epsilon(p+k) - \Delta) \\
\label{eq40} \Delta = 2IN^{-1/2}<S^{z}_{0}> - U (n_{\uparrow} -
n_{\downarrow}) = 2 I \bar S - U m =   \Delta_{I} +
\Delta_{U}  \\
\chi ^{s}_{0}(k,\omega) = N^{-1} \sum_{p} \frac {
(f_{p+k\downarrow} - f_{p\uparrow})}{\omega_{p,k}}
\end{eqnarray}
Now we consider the GF $<<\sigma^{+}_{k}(t),B(t')>>$ . Similarly
to Eq.( \ref{eq30}), we have
\begin{eqnarray} \label{eq41}
-N^{1/2}I\chi ^{s}_{0}(k,\omega)<<S^{+}_{k} \vert B
>>_{\omega} +  ( 1 - U\chi ^{s}_{0}(k,\omega))<<\sigma^{+}_{k} \vert
B>>_{\omega} =   \nonumber \\    {0 \brace -N \chi
^{s}_{0}(k,\omega) } + \sum_{p} \frac {1}{\omega_{p,k}} <<A_{2}
\vert B>>_{\omega}
\end{eqnarray}
Here the following definition of the irreducible part for the
Coulomb correlation term was used
\begin{eqnarray} \label{eq42}
(a^{\dagger}_{p\uparrow}a^{\dagger}_{q+r\uparrow}a_{q\uparrow}a_{p+r+k\downarrow}
-
a^{\dagger}_{p+r\uparrow}a^{\dagger}_{q-r\downarrow}a_{q\downarrow}a_{p+k\downarrow})^{~ir}
\nonumber \\ =
(a^{\dagger}_{p\uparrow}a^{\dagger}_{q+r\uparrow}a_{q\uparrow}a_{p+r+k\downarrow}
-
a^{\dagger}_{p+r\uparrow}a^{\dagger}_{q-r\downarrow}a_{q\downarrow}a_{p+k\downarrow})
\nonumber \\
- <a^{\dagger}_{q+r\uparrow}a_{q\uparrow}>\delta_{q+r,q
}a^{\dagger}_{p\uparrow}a_{p+r+k\downarrow} \nonumber \\ -
<a^{\dagger}_{q-r\downarrow}a_{q\downarrow}> \delta_{q-r,q
}a^{\dagger}_{p+r\uparrow}a_{p+k\downarrow}
\end{eqnarray}
The operator $A_{2}$   satisfies the conditions
$$ <[A_{2},S^{-}_{-k}]_{-}> = <[A_{2},\sigma^{-}_{-k}]_{-}> = 0$$
In the matrix notation  the full equation of motion for the GF
$\hat G(k;\omega)$ can now be summarized   in the following form:
\begin{eqnarray}
\label{eq43} \hat \Omega \hat G(k;\omega) = \hat I + \sum_{p}\hat
\Phi(p) \hat D(p;\omega) \\
(\hat G(k;\omega))^{\dagger} (\hat \Omega)^{\dagger}  = (\hat
I)^{\dagger} + \sum_{p} (\hat D(p;\omega))^{\dagger} (\hat
\Phi(p))^{\dagger}  \nonumber
\end{eqnarray}
where
\begin{eqnarray}
\label{eq44} \quad \hat \Omega = \pmatrix{ \Omega_{1} &
\Omega_{2}\cr -IN^{1/2}\chi^{s}_{0}&(1 - U\chi^{s}_{0} ) \cr}
\quad  \hat I = \pmatrix{ I^{-1}N^{1/2}\Omega_{2} & 0\cr
0&-N\chi^{s}_{0}\cr} \\ \hat D(p;\omega) =  \pmatrix{ <<A_{1}
\vert S^{-}_{k}>> & <<A_{1} \vert \sigma^{-}_{-k}>> \cr <<A_{2}
\vert S^{-}_{-k}>> & <<A_{2} \vert \sigma^{-}_{-k}>> \cr} \quad
\hat \Phi(p) = \pmatrix{ N^{-1} & 0\cr 0&\omega^{-1}_{p,k}\cr}
\end{eqnarray}
To calculate the higher order GFs in ( \ref{eq43}), we
differentiate its r.h.s.   with respect to the second-time
variable (t'). Let us give explicitly one of the four equations.
After introducing the irreducible parts as discussed above we get
\begin{eqnarray}\label{eq45}
<< A \vert S^{-}_{-k} >>_{\omega} \Omega_{1} = \frac {I}{N}
\sum_{p'q'} << A \vert
S^{-}_{-(k-q')}(a^{\dagger}_{p'\uparrow}a_{p'+q'\uparrow} -
a^{\dagger}_{p'\downarrow}a_{p'+q'\downarrow})^{~ir} -
2S^{z}_{-(k-q')}a^{\dagger}_{p'\downarrow}a_{p'+q'\uparrow}
>>_{\omega} \nonumber\\  + N^{-1/2}\sum_{q'} J_{q'}<<A \vert [(
S^{z}_{q'})^{~ir} S^{-}_{-(k+q')}  - (S^{z}_{-(k+q')})^{~ir}
S^{-}_{q'}]^{~ir}>>_{\omega}
\end{eqnarray}
Here, the quantity A in the l.h.s.   should be substituted by
\[ A = \left\{
\begin{array}{rl}
( (S^{z}_{q})~^{ir}S^{+ }_{k-q}  -
(S^{z}_{k-q})~^{ir}S^{+}_{q})~^{ir} \\
S^{+}_{k-q}(a^{\dagger}_{p\uparrow}a_{p+q\uparrow} -
a^{\dagger}_{p\downarrow}a_{p+q\downarrow})~^{ir}  \\
2S^{z}_{k-q}a^{\dagger}_{p\uparrow}a_{p+q\downarrow}
\end{array} \right. \]
In the matrix notation  the full equation of motion for the GF
$\hat D(k;\omega)$ can now be written   in the following form:
\begin{equation}
\label{eq46} \hat \Omega \hat D(p;\omega) =   \sum_{p'}\hat \Phi
(p') \hat D_{1}(p';\omega)
\end{equation}
where
\begin{equation} \label{eq47}
\hat D_{1}  =  \pmatrix{ <<A_{1} \vert A^{\dagger}_{1}>> &
<<A_{1} \vert A^{\dagger}_{2}>> \cr <<A_{2} \vert
A^{\dagger}_{1}>> & <<A_{2} \vert A^{\dagger}_{2} >> \cr}
\end{equation}
Combining both (the first- and second-time differentiated)
equations of motion, we get the "exact"( no approximation has
been made till now) "scattering" equation
\begin{equation} \label{eq48}
\hat \Omega \hat G(k;\omega) = \hat I +  \sum_{pp'}\hat \Phi(p)
\hat P(p,p') \hat \Phi(p') (\hat \Omega^{\dagger})^{-1}
\end{equation}
This equation can be identically transformed to the standard form
Eq.( \ref{eq13})
\begin{eqnarray}\label{eq49}
\hat G = \hat G_{0} + \hat G_{0} ( \sum_{pp'} \hat I^{-1} \Phi(p)
\hat P(p,p')  \Phi(p') \hat I^{-1}) \hat G_{0}\nonumber \\
\label{eq50} \hat G = \hat G_{0} + \hat G_{0} \hat P \hat G_{0}
\end{eqnarray}
Here we have introduced the generalized mean-field (GMF) GF
$G_{0}$, according to the following definition:
\begin{equation}
\label{eq51} \hat G_{0} = \hat \Omega^{-1} \hat I
\end{equation}
The scattering operator $P$ has the form
\begin{equation}
\label{eq52} \hat P =
  \hat I^{-1} \sum_{pp'} \hat \Phi(p) \hat P(p,p') \hat \Phi(p') \hat
I^{-1}
\end{equation}
Here we have used the obvious notation
\begin{equation} \label{eq53}
\hat P(p,p';\omega) = \pmatrix{ <<A_{1}  \vert A^{\dagger}_{1} >>
& <<A_{1}  \vert A^{\dagger}_{2}
>> \cr <<A_{2}  \vert A^{\dagger}_{1} >> & <<A_{2}
\vert A^{\dagger}_{2} >> \cr}
\end{equation}
As is shown above, Eq.( \ref{eq49}) can be transformed exactly
into the Dyson equation ( \ref{eq15})
\begin{equation}
\label{eq54} \hat G = \hat G_{0} + \hat G_{0} \hat M\hat G_{0}
\end{equation}
with the self-energy operator $M$ given as
\begin{equation}
\label{eq55} \hat M  = (\hat P)^{p}
\end{equation}
Hence, the determination of the full GF $\hat G$ has been reduced
to that of $\hat G_{0}$ and $\hat M$.
\section{Generalized Mean-Field GF }
From the definition ( \ref{eq51}), the GF matrix in the
generalized mean-field approximation reads
\begin{equation} \label{eq56}
\hat G_{0} = R^{-1} \pmatrix{ (1 - U\chi^{s}_{0} )
I^{-1}N^{1/2}\Omega_{2} & \Omega _{2} N \chi^{s}_{0}\cr \Omega
_{2} N \chi^{s}_{0} &-\Omega_{1}N\chi^{s}_{0}\cr}
\end{equation}
where
$$R =
(1 - U\chi^{s}_{0} )\Omega_{1}
+ \Omega _{2}I N^{1/2} \chi^{s}_{0}$$ \\
Let us write down explicitly the diagonal matrix elements
$G^{11}_{0}$ and $G^{22}_{0}$
\begin{equation} \label{eq57}
<<S^{+}_{k} \vert S^{-}_{-k}>>^{0} = \frac {2 \bar S}{ \Omega_{1}
+ 2I^{2} \bar S \chi^{s}(k,\omega ) }
\end{equation}
\begin{equation} \label{eq58}
<<\sigma^{+}_{k} \vert \sigma^{-}_{-k}>>^{0} = \frac {  \Omega_{1}
\chi^{s} (k,\omega ) }{\Omega_{1} + 2I^{2}\bar S \chi^{s}
(k,\omega )}
\end{equation}
where
\begin{eqnarray} \label{eq59}
\chi^{s}(k,\omega ) = \chi^{s}_{0}(k,\omega )(1 -
U\chi^{s}_{0}(k,\omega ) )^{-1}\\ \nonumber
\bar S = N^{-1/2}<S^{z}_{0}>
\end{eqnarray}
To clarify the functional structure of the generalized mean-field
GFs  (\ref{eq57} ) and (\ref{eq58} ), let us consider a few
limiting cases.
\subsection{Uncoupled Subsystems}
To clarify the calculation of quasiparticle spectra of coupled
localized and itinerant subsystems, it is instructive to consider
an artificial limit of uncoupled subsystems.  We then assume that
the local exchange parameter I = 0. In this limiting case we have
\begin{equation} \label{eq60}
<<S^{+}_{k} \vert S^{-}_{-k}>>^{0} = \frac {2 \bar S}{   \omega -
\bar S ( J_{0} - J_{k} ) -  \frac {1}{2N \bar S} \sum_{q}( J_{q} -
J_{q-k} )(2K^{zz}_{q} + K^{-+}_{q})  }
\end{equation}
\begin{equation} \label{eq61}
<<\sigma^{+}_{k} \vert \sigma^{-}_{-k}>>^{0} = \chi^{s} (k,\omega
)
\end{equation}
 The spectrum of quasi-particle
excitations of localized spins without damping follows from the
poles of the generalized mean-field GF (\ref{eq60})
\begin{equation} \label{eq62}
    \omega(k) =
\bar S ( J_{0} - J_{k} ) +  \frac {1}{2N \bar S} \sum_{q}( J_{q} -
J_{q-k} )(2K^{zz}_{q} + K^{-+}_{q})
\end{equation}
It is seen that due to the correct definition of generalized mean
fields we get the result for the localized spin Heisenberg
subsystem which includes both the simplest spin-wave result and
the result of Tyablikov decoupling as  limiting cases. In the
hydrodynamic limit $k \rightarrow 0$, $\omega \rightarrow 0$   it
leads to the dispersion law $\omega(k) = Dk^{2}$.\\ The exchange
integral $J_{k}$ can be written in the following way:
\begin{equation} \label{eq63}
J_{k} = \sum_{i} \exp { (-i \vec k \vec R_{i})} J( |\vec R_{i}|)
\end{equation}
The expansion in small $\vec k $ gives
\begin{equation} \label{eq64}
J_{k} = \sum_{i} J( |\vec R_{i}|) - \frac{1}{2} \sum_{i} (\vec k
 \vec R_{i})^{2} J( |\vec R_{i}|)
 = J_{0} - \frac{k^2}{2} \sum_{i} (\vec n
 \vec R_{i})^{2} J( |\vec R_{i}|)
\end{equation}
Here $\vec n = \vec k/k $ is the unit vector. The values $J_{k-q}
$ can be evaluated in a similar way
\begin{eqnarray} \label{eq65}
J_{k-q} = J_{q}  - ( \vec k \nabla_{q} )J_{q} + \frac{1}{2}( \vec
k \nabla_{q} )^{2} J_{q} +  \cdots \\ \nonumber ( \vec k
\nabla_{q} )J_{q} = - i   \sum_{i} (\vec k
 \vec R_{i})  J( |\vec R_{i}|)\exp { (-i \vec q \vec R_{i})}\\
 \nonumber
( \vec k \nabla_{q} )^{2} J_{q} = -  \frac{1}{2}   \sum_{i} (\vec
k \vec R_{i})^{2} J( |\vec R_{i}|)\exp { (-i \vec q \vec R_{i})}
\end{eqnarray}
Combining Eq.(\ref{eq65}), Eq.(\ref{eq64}), and Eq.(\ref{eq62}) we
get
\begin{eqnarray} \label{eq66}
<<S^{+}_{k} \vert S^{-}_{-k}>>^{0} = \frac {2  \bar
 S }{ \omega - \omega(k)} \\ \nonumber
    \omega(k\rightarrow 0) =
\Bigl ( \bar S ( J_{0} - J_{k} ) +  \frac {1}{2N \bar S} \sum_{q}(
J_{q} - J_{q-k} )(2K^{zz}_{q} + K^{-+}_{q})\Bigr )   \simeq D_{1}k^{2}\\
\nonumber  = \Bigl (  \frac {\bar S}{2} \psi_{0} + \frac{N}{2\bar
S^{2}}\sum_{q} \psi_{q}(2K^{zz}_{q} + K^{-+}_{q})  \Bigr ) k^{2}\\
\nonumber \psi_{q} =  \sum_{i} (\vec k \vec R_{i})^{2} J( |\vec
R_{i}|)\exp { (-i \vec q \vec R_{i})}
\end{eqnarray}
Let us now consider   the spin susceptibility of itinerant
carriers Eq. (\ref{eq61} ) in the hydrodynamic limit $k
\rightarrow 0$, $\omega \rightarrow 0$. It is convenient to
consider the static limit of Eq. (\ref{eq61} )
\begin{eqnarray} \label{eq67}
<<\sigma^{+}_{k} \vert \sigma^{-}_{-k}>>^{0}|_{\omega = 0} =
\frac{\chi^{s}_{0} (k,0 )}{1 - U \chi^{s}_{0} (k,0)}\\ \nonumber
\chi^{s}_{0} (k,0 ) = \frac{1}{N}\sum_{q} \frac{
f_{q+k\downarrow} - f_{q\uparrow}}{  \epsilon(q) -\epsilon(q+k) -
\Delta_{U} } \\ \nonumber \Delta_{U}    =   U (n_{\uparrow} -
n_{\downarrow}) = U m
\end{eqnarray}
To proceed, we make a small-k expansion of the form
\begin{eqnarray}\label{eq68}
\epsilon(q+k) -\epsilon(q) = ( \vec k \nabla_{q} )\epsilon(q) +
\frac{1}{2}( \vec k \nabla_{q} )^{2}\epsilon(q) + \cdots \\
\nonumber \chi^{s}_{0} (k,0) = \frac{1}{N \Delta_{U}}\sum_{q} (
f_{q\uparrow} - f_{q\downarrow}) - \\ \nonumber \frac{1}{N
\Delta_{U}^2} \sum_{q}  ( f_{q\uparrow} + f_{q\downarrow})
\frac{1}{2}( \vec k \nabla_{q} )^{2}\epsilon(q) + \frac{1}{N
\Delta_{U}^3}\sum_{q} ( f_{q\uparrow} - f_{q\downarrow}) ( \vec k
\nabla_{q} \epsilon(q))^{2} + \cdots
\end{eqnarray}
The   poles of the spin susceptibility of itinerant carriers are
determined by the equation
\begin{equation} \label{eq69}
1 - U \chi^{s}_{0} (k,\omega) = 0
\end{equation}
In another form this reads in detail
$$ 1 = \frac{U}{N} \sum_{q} \frac{f_{q\uparrow} - f_{q+k\downarrow}}{
\epsilon(k+q) - \epsilon(q) + \Delta_{U} - \omega}$$ If we set
$\omega = E(k)$  and put then $ k = 0$, we get the equation for
the excitation energy $E(k=0)$
$$ 1 = \frac{U}{N} \sum_{q} \frac{ f_{q\uparrow} - f_{q\downarrow} }{
  \Delta_{U} - E(k=0) } = \frac{U}{\Delta_{U} - E(k=0)} \frac{\Delta_{U}}{U}$$
which is satisfied if $E(k=0) = 0$. Thus, a solution of
Eq.(\ref{eq69} ) exists which has the property
$\lim_{k\rightarrow 0} E(k) = 0$ and this solution corresponds to
an acoustic spin-wave branch of excitations
\begin{eqnarray} \label{eq70}
E(k) = D_{2}k^{2} = - \frac {U}{2N \Delta_{U}}
\sum_{q}(f_{q\uparrow} + f_{q\downarrow})(\vec k  \nabla_{q})^{2}
\epsilon(\vec q) +   \frac {U}{N \Delta_{U}^2 }
\sum_{q}(f_{q\uparrow}  - f_{q\downarrow})(\vec k \nabla_{q}
\epsilon(\vec q) )^{2} \\
\nonumber \omega = \epsilon(k+q) - \epsilon(q) +  \Delta_{U}
\end{eqnarray}
It is seen that the stiffness constant $D_{2}$ can be interpreted
as expanded in $ \frac{1}{\Delta_{U} }$.  For the tight-binding
electrons in s.c. lattice the spin wave dispersion relation $
D_{2}k^{2}$ becomes
\begin{eqnarray} \label{eq71}
 D_{2} k^{2} =   ( 3(n_{\uparrow} -
n_{\downarrow}))^{-1} \sum_{q}[\frac{(f_{q\uparrow}  -
f_{q\downarrow})}{{\Delta_{U} }} |\nabla_{q} \epsilon(\vec
q)|^{2}   -    \frac{(f_{q\uparrow} + f_{q\downarrow})}{2}
\nabla_{q}^{2} \epsilon(\vec q)] =
\\ \nonumber (
3(n_{\uparrow} - n_{\downarrow}))^{-1}  ( \frac
{2t^{2}a^{2}}{\Delta_{U} } \sum_{q} (f_{q\uparrow} -
f_{q\downarrow})( k_{x} \sin (q_{x}a) + k_{y} \sin (q_{y}a) +
k_{z} \sin (q_{z}a))^{2} -
\nonumber \\
ta^{2} \sum_{q} (f_{q\uparrow} + f_{q\downarrow})(k^{2}_{x} \cos
q_{x}a + k^{2}_{y} \cos q_{y}a + k^{2}_{z} \cos q_{z}a))\nonumber
\end{eqnarray}
\subsection{Coupled Subsystems}
The next stage in the analysis of the quasi-particle spectra of
the $(sp-d)$ model is the introduction of the nonzero coupling I.
The full generalized mean field GFs can be rewritten as
\begin{equation} \label{eq72}
<<S^{+}_{k} \vert S^{-}_{-k}>>^{0} = \frac {2 \bar S}{   \omega -
Im - \bar S ( J_{0} - J_{k} ) -  \frac {1}{2N \bar S} \sum_{q}(
J_{q} - J_{q-k} )(2K^{zz}_{q} + K^{-+}_{q}) + 2I^{2} \bar S
\chi^{s} (k,\omega)}
\end{equation}
\begin{equation} \label{eq73}
<<\sigma^{+}_{k} \vert \sigma^{-}_{-k}>>^{0} = \frac{\chi^{s}_{0}
(k,\omega )}{1 - U_{eff}(\omega) \chi^{s}_{0} (k,\omega)}
\end{equation}
Here the notation is used
$$ U_{eff} = U - \frac{2I^{2} \bar S}{\omega - Im }; \quad m =
( n_{\uparrow} - n_{\downarrow}) $$
The expression Eq.(\ref{eq73}) coincides with that for the
itinerant spin susceptibility as calculated in\cite{don}.
It is instructive to consider separately the four different cases,
\begin{itemize}
\item [(i)]$I \neq 0, J = 0, U = 0$
\item [(ii)] $I \neq 0, J \neq 0, U = 0$
\item [(iii)]$I \neq 0, J = 0, U \neq 0$
\item [(iv)]$I \neq 0, J \neq 0, U  \neq 0$
\end{itemize}
\subsubsection{}
The first case
 $I \neq 0, J = 0, U = 0$ corresponds to a model
which is commonly called the Kondo lattice model.    It can be
seen that GFs (\ref{eq72} ) and (\ref{eq73} )  are then equal to
\begin{equation} \label{eq74}
<<S^{+}_{k} \vert S^{-}_{-k}>>^{0} = \frac {2 \bar S}{   \omega -
Im   + 2I^{2} \bar S \chi_{0}^{s} (k,\omega)}
\end{equation}
\begin{equation} \label{eq75}
<<\sigma^{+}_{k} \vert \sigma^{-}_{-k}>>^{0} = \frac{\chi^{s}_{0}
(k,\omega )}{ \omega  +  \frac {2 I^{2} \bar S}{\omega - Im }
\chi^{s}_{0} (k,\omega)}
\end{equation}
In order to calculate the acoustic  pole of the GF  (\ref{eq74} ),
we make use of the small $(k,\omega)$ expansion. Hence we get
\begin{eqnarray} \label{eq76}
<<S^{+}_{k} \vert S^{-}_{-k}>>^{0}  \approx
\\ \nonumber \frac {2 \bar S (1 +
\frac{m}{2 \bar S })^{-1} }{ \omega - (1 + \frac{m}{2 \bar S
})^{-1} [\frac {1}{2N \Delta_{I}^2} \sum_{q}(f_{q\uparrow} +
f_{q\downarrow})(\vec k \nabla_{q})^{2} \epsilon(\vec q) - \frac
{1}{N \Delta_{I}^3 } \sum_{q}(f_{q\uparrow}  -
f_{q\downarrow})(\vec k \nabla_{q} \epsilon(\vec q) )^{2} ]}
\end{eqnarray}
It follows from   Eq.(\ref{eq76} ) that the stiffness constant D
is proportional to the total magnetization of the system.
\subsubsection{}
In the second case  $I \neq 0, J \neq 0, U = 0$, we get
\begin{equation} \label{eq77}
<<S^{+}_{k} \vert S^{-}_{-k}>>^{0} = \frac {2 \bar S}{   \omega -
Im - \bar S ( J_{0} - J_{k} ) -  \frac {1}{2N \bar S} \sum_{q}(
J_{q} - J_{q-k} )(2K^{zz}_{q} + K^{-+}_{q}) + 2I^{2} \bar S
\chi_{0}^{s} (k,\omega)}
\end{equation}
\begin{equation} \label{eq78}
<<\sigma^{+}_{k} \vert \sigma^{-}_{-k}>>^{0} = \frac{\chi^{s}_{0}
(k,\omega )}{1 -  \frac{2I^{2} \bar S}{\omega - Im } \chi^{s}_{0}
(k,\omega)}
\end{equation}
In order to calculate the acoustic  pole of the GF  (\ref{eq77} ),
we make use of the small $(k,\omega)$ expansion again. We then get
\begin{eqnarray} \label{eq79}
<<S^{+}_{k} \vert S^{-}_{-k}>>^{0}  \approx
\\ \nonumber \frac {2 \bar S (1 +
\frac{m}{2 \bar S })^{-1} }{ \omega - (1 + \frac{m}{2 \bar S
})^{-1}
 D_{1}k^{2} - (1 + \frac{m}{2 \bar S })^{-1} [\frac {1}{2N
\Delta_{I}^2} \sum_{q}(f_{q\uparrow} + f_{q\downarrow})(\vec k
\nabla_{q})^{2} \epsilon(\vec q) -   \frac {1}{N \Delta_{I}^3 }
\sum_{q}(f_{q\uparrow}  - f_{q\downarrow})(\vec k \nabla_{q}
\epsilon(\vec q) )^{2} ]}
\end{eqnarray}
It follows from   Eqs.(\ref{eq76} ) and  (\ref{eq79} ) that the
stiffness constant D is proportional to the total magnetization
of the system.
\subsubsection{}
The third case  $I \neq 0, J = 0, U \neq 0$ corresponds to a model
which is  called the modified Zener lattice model\cite{bar}.    It
can be seen that  in this case  GFs (\ref{eq72} ) and (\ref{eq73}
)   are equal
 to
\begin{equation} \label{eq80}
<<S^{+}_{k} \vert S^{-}_{-k}>>^{0} = \frac {2 \bar S}{   \omega -
Im   + 2I^{2} \bar S \chi^{s} (k,\omega)}
\end{equation}
\begin{equation} \label{eq81}
<<\sigma^{+}_{k} \vert \sigma^{-}_{-k}>>^{0} = \frac{\chi^{s}_{0}
(k,\omega )}{1 - U_{eff}(\omega) \chi^{s}_{0} (k,\omega)}
\end{equation}
The results obtained here coincide with those of Bartel\cite{bar}.
The excitation energies for the localized spin and spin densities
of itinerant carriers are found from the zeros of the
denominators of $<<S^{+}_{k} \vert S^{-}_{-k}>>^{0} $ and
$<<\sigma^{+}_{k} \vert \sigma^{-}_{-k}>>^{0} $ which yield
identical excitation spectra, consisting of three branches, the
acoustic  spin wave $E^{ac}(k)$, the optical spin wave
$E^{op}(k)$, and the Stoner continuum $E^{St}(k)$
$$E^{ac}(k) =Dk^{2}$$
$$E^{op}(k) = E^{op}_{0} - D (1 - \frac {U E^{op}}{I \Delta}) k^{2};
\quad  E^{op}_{0} =  I ( m + 2 \bar S )
   $$
$$E^{St}(k) =  \epsilon(k+q) - \epsilon(q) + \Delta$$
\subsubsection{}
The most general is the forth case, $I \neq 0, J \neq 0, U \neq
0$. The total GF of the coupled system is given by Eq.(\ref{eq72}
). The magnetic excitation spectrum follows from the poles of the
GF ( \ref{eq56})
$$R =
(1 - U\chi^{s}_{0} )\Omega_{1} + \Omega _{2}I N^{1/2}
\chi^{s}_{0} = 0 $$ and consists of three branches - the acoustic
spin wave $E^{ac}(k)$, the optical spin wave
$E^{op}(k)$, and the Stoner continuum $E^{St}(k)$.\\
Let us, as a first approximation, consider the last term in its
denominator , which is the dynamic spin susceptibility of
itinerant carriers, in the static limit, without any frequency
dependence. The GF Eq.(\ref{eq72} ) then becomes equal to
\begin{equation} \label{eq82}
<<S^{+}_{k} \vert S^{-}_{-k}>>^{0}  \approx \frac {2 \bar S}{
\omega - Im - \bar S ( J_{0} - J_{k} ) -  \frac {1}{2N \bar S}
\sum_{q}( J_{q} - J_{q-k} )(2K^{zz}_{q} + K^{-+}_{q}) + 2I^{2}
\bar S \chi^{s} (k,0)}
\end{equation}
It is possible to verify that in the limit $k \rightarrow 0$
\begin{equation} \label{eq83}
2I^{2} \bar S \chi^{s} (k,0)  \approx  Im -  \frac {1}{2 \bar S N}
\sum_{q}(f_{q\uparrow} + f_{q\downarrow})(\vec k \nabla_{q})^{2}
\epsilon(\vec q) + \frac {1}{2 \bar S N \Delta}
\sum_{q}(f_{q\uparrow} - f_{q\downarrow})(\vec k \nabla_{q}
\epsilon(\vec q) )^{2}
\end{equation}
Then for $\omega, k \rightarrow 0$  Eq.(\ref{eq83} ) becomes
\begin{eqnarray} \label{eq84}
<<S^{+}_{k} \vert S^{-}_{-k}>>^{0}  \approx
\\ \nonumber \frac {2 \bar S}{ \omega - D_{1}k^{2} -
\frac {1}{2 \bar S 2N} \sum_{q}(f_{q\uparrow} +
f_{q\downarrow})(\vec k \nabla_{q})^{2} \epsilon(\vec q) + \frac
{1}{2 \bar S N \Delta} \sum_{q}(f_{q\uparrow} -
f_{q\downarrow})(\vec k \nabla_{q} \epsilon(\vec q) )^{2}}
\end{eqnarray}
This expression can be expected to be qualitatively correct in
spite of the primitive approximation.
The spectrum of Stoner excitations is given by
\begin{equation} \label{eq85}
 E^{St}(k) = \epsilon(k+q) - \epsilon(q) + \Delta
\end{equation}
In addition to the acoustic  branch there is an optical branch of
spin excitations.  This can be seen from the following: For $ k =
0$ we get for R = 0 the quadratic equation in $\omega$ with two
solutions, $ \omega = 0$ and $ \omega = I ( m + 2 \bar S ) =
E^{op}_{0}  $. In the hydrodynamic limit, $k \rightarrow 0$,
$\omega \rightarrow 0$ the GF (\ref{eq71}) can be written as
\begin{equation} \label{eq86}
<<S^{+}_{k} \vert S^{-}_{-k}>>^{0}   \simeq   \frac {2 \bar S}{
\omega - E^{ac}(k)}
\end{equation}
where the acoustic  spin wave energies are given by
\begin{eqnarray} \label{eq87}
E^{ac}(k) = Dk^{2} = \Bigl (  \frac {\bar S}{2} [\psi_{0} +
\frac{1}{2N \bar S^{2}}\sum_{q} \psi_{q}(2K^{zz}_{q} +
K^{-+}_{q})]\\ \nonumber + \frac {1}{2N} \frac{1}{2 \bar S}
\sum_{q}(f_{q\uparrow} + f_{q\downarrow})(\vec n \nabla_{q})^{2}
\epsilon(\vec q) + \frac {1}{N  \Delta} \frac{1}{2 \bar S}
\sum_{q}(f_{q\uparrow}  - f_{q\downarrow})(\vec n \nabla_{q}
\epsilon(\vec q) )^{2}  \Bigr ) k^{2}
\end{eqnarray}
For the optical spin wave branch the estimations can be carried
out as in paper\cite{don}
\begin{equation} \label{eq88}
E^{op}(k) = E^{op}_{0} - D^{op} k^{2}
\end{equation}
In the GMF approximation the density of itinerant electrons ( and
the band splitting $\Delta$) can be evaluated by solving the
equation
\begin{equation} \label{eq89}
n_{\sigma} =   \frac{1}{N}  \sum_{k} [\exp (\beta(\epsilon(k) +
Un_{-\sigma} - I \bar S - \epsilon_{F})) + 1]^{-1}
\end{equation}
Hence, the stiffness constant $D$ can be expressed by the
parameters of the $sp-d$ model Hamiltonian  .
\section{ Effects of Disorder in DMS}
We now proceed to give a simple and qualitative discussion of the
effects of disorder in DMS.  The main aim of the investigation of
DMS is to give a successful microscopic picture of the
ferromagnetic ordering of localized spins induced by the
interaction with the spin density of itinerant charge carriers.
The big amount of information was accumulated on the problem of
ordering of localized spins of $Mn^{2+}$ ions in
$Cd_{1-x}Mn_{x}Te$ and similar compounds\cite{di1},\cite{tak9}.
Recently, main attention was directed to studying III-V
semiconducting alloys doped with Mn. As it was stated above, a
suitable model, which may be used for  investigation of this
problem ( at least at the initial stage ) is a  modified Kondo
lattice model (\ref{eq5c} )
\begin{equation}
\label{eq90} H  = \sum_{ij} \sum_{\sigma}
t_{ij}a^{\dagger}_{i\sigma}a_{j\sigma} - \sum_{i}2I \nu_{i} {\vec
\sigma_{i}}{\vec S_{i}}
\end{equation}
Here $\nu_{i} $ projects out sites occupied by Mn atoms, i.e.:
$$\nu_{i}  = \cases{ 1 & \text{if site i is occupied by Mn}  \cr
0 & \text{if  site i is occupied by Ga}   \cr} $$
This model is relevant for the doped II-VI or III-V compound. The
essential feature of the model is that it describes a mechanism
of how the spins of carriers (electrons or holes ) become
polarized due to the local antiferromagnetic exchange
interactions with localized spins. In $A^{III}_{1-x}Mn_{x}B^{V} $
the main magnetic interaction is an antiferromagnetic exchange
between the Mn spins and the charge-carrier spins. The
superexchange term $H_{d} = -\frac{1}{2} \sum_{ij} J_{ij} \vec
S_{i}\vec S_{j} $ is antiferromagnetic also but is as a rule
rather small in the concentration range of interest ( $ x \approx
0.05$). In the case of Mn-doped III-V compounds the
antiferromagnetic superexchange interaction will generally reduce
the ferromagnetic ordering temperature. As a result, the
carrier-induced ferromagnetism in DMS may arise   due to the
effective ferromagnetic interaction between the Mn spins. In
other words, the ferromagnetism in this system is most probably
related to the uncompensated Mn spins and is mediated by holes.
The density of Mn ions $c_{Mn}$ is greater than the hole density
$p$, $c_{Mn} \gg p$. The optimal interrelation of both the
magnitudes is a delicate and subtle question and was analyzed
recently in paper\cite{yu}. It was shown that the concentration
of free holes and ferromagnetically active Mn spins were governed
by the position of the Fermi level  which controls the formation
energy of compensating interstitial Mn donors. The experimental
evidence has been provided that the upper limit of the Curie
temperature is caused by Fermi-level-induced hole saturation. In
order to provide a suitable treatment of the spin quasiparticle
dynamics it is necessary to take into account the effects of
disorder since the Mn ions are assumed to be distributed randomly
with concentration $c$. This is positional disorder. There is
variation of site-energy of nonmagnetic origin due to the
substitution of A atom with Mn ion. The detailed nature of the
disorder is not fully clear. In paper\cite{yu}, it was shown that
the dominant fraction of the Mn atoms are on either
substitutional sites or   specific sites shadowed by the host
atoms. This reveals that the majority of the Mn atoms are on
specific ( nonrandom ) sites commensurate with the lattice, but
that does not necessarily imply that all of the Mn atoms are in
substitutional positions. For $ x > 0.05$  an increasing fraction
of Mn spins do not participate in ferromagnetism. It can be
related with an increase in the concentration of Mn interstitials
accompanied by a reduction of $T_{c}$. There are indications of
an increase in Mn atoms in the form of random clusters not
commensurate with the
GaAs lattice.\\
It follows from   Eq.(\ref{eq90} )     that the spin dynamics of
a modified KLM will be described by the GFs in the lattice site
representation for a given configuration
$$<<S^{+}_{i} \vert S^{-}_{j}>> \quad
<<\sigma^{+}_{i} \vert \sigma^{-}_{j}>>$$
and instead of Eq.(\ref{eq25}) the lattice GF should be considered
\begin{equation}\label{eq25a}
\pmatrix{ <<S^{+}_{i}\vert S^{-}_{j}>> & <<S^{+}_{i}\vert
\sigma^{-}_{j}>> \cr <<\sigma^{+}_{i}\vert S^{-}_{j}>> &
<<\sigma^{+}_{i}\vert \sigma^{-}_{j}>> \cr} = \hat G_{ij}( \omega)
\end{equation}
In order to provide a simultaneous and self-consistent treatment
of the quasi-particle dynamics including the effects of disorder,
a sophisticated description of disorder should be done. Most
treatments remove disorder by making a virtual-crystal-like
approximation in which the Mn ion distribution is replaced by a
continuum. A more sophisticated approach for treating the
positional disorder of the magnetic impurities inside the host
semiconductor is the CPA\cite{tak6}-\cite{tak8}. The CPA replaces
the initial Hamiltonian of disordered system by an effective one
which is assumed to produce no further scattering\cite{sov}. It
describes reasonably well the state of itinerant charge
scattering in disordered substitutional alloys $A_{1-x}B_{x}$.\\
In order to simplify the discusson     here, we will deal with a
much simpler and less sophisticated description. The
approximation discussed below should be considered as a first,
crude approximation to a theory of disorder effects in DMS.  Since
the detailed nature of disorder in DMS is not yet   established
completely, we will confine ourselves to the simplest possible
approximation. Let us remind that the IGF method is based on the
suitable definition of the $ \emph{ generalized   mean  fields }
$\cite{kuz2}.  To demonstrate the flexibility of the IGF method,
we   show below how the mean field should be redefined to include
the disorder in an effective way. The previous definition of the
irreducible spin operator, Eq.( \ref{eq28}), should be replaced by
\begin{eqnarray}\label{eq91}
(S^{z}_{q})~^{ir} = S^{z}_{q} - c \overline{ <S_{z}>}
\delta_{q,0}; \quad (a^{\dagger}_{p+q\sigma}a_{p\sigma})~^{ir} =
a^{\dagger}_{p+q\sigma}a_{p\sigma} -
<a^{\dagger}_{p\sigma}a_{p\sigma}>\delta_{q,0}
\end{eqnarray}
Here $ \overline {<S_{z}} > = \bar S_{z}$ corresponds   to the
configuration average.  The average $<S_{z}>$ denotes the mean
value of $S^{z}$ for a given configuration of all   the spins. We
omitted here the variation of site energy of nonmagnetic origin.
The consequences of this choice manifest  themselves. It means
precisely that in a random system the mean field is weaker as
compared to a regular system. The approximation is conceptually as
simple as an ordinary mean field approximation and corresponds to
the virtual crystal approximation. The situation is then
completely analogous to the previous one considered in the
preceding sections. We get for the configurationally averaged GFs
\begin{equation} \label{eq92}
\overline {<<S^{+}_{i} \vert S^{-}_{j}>>^{0}} = <<S^{+}_{k} \vert
S^{-}_{-k}>>^{0} \approx  \frac {2 c \bar S_{z}}{ \omega - Im   +
2I^{2} c \bar S_{z} \chi_{0}^{s} (k,\omega)}
\end{equation}
\begin{equation} \label{eq93}
\overline {<<\sigma^{+}_{i} \vert \sigma^{-}_{j}>>^{0}} =
<<\sigma^{+}_{k} \vert \sigma^{-}_{-k}>>^{0} \approx
\frac{\chi^{s}_{0} (k,\omega )}{ \omega  +  \frac {2 I^{2} c \bar
S_{z}}{\omega - Im } \chi^{s}_{0} (k,\omega)}
\end{equation}
Here $ \bar S_{z} = N^{-1/2} \overline{ <S_{z}>}$.
These simple results are  fully tractable and are the reason for
their derivation.\\
It is worth to note that in the case of the modified Zener model
which contains the correlation (Hubbard) term, the effects of
disorder should be considered on the basis of a similar
model\cite{ku}
\begin{equation}
\label{eq94} H  = \sum_{ij} \sum_{\sigma}
t_{ij}a^{\dagger}_{i\sigma}a_{j\sigma} + U \sum_{i}  \nu_{i}
n_{i\uparrow}  n_{i\downarrow}  - \sum_{i}2I \nu_{i} {\vec
\sigma_{i}}{\vec S_{i}}
\end{equation}
The Coulomb repulsion is assumed to exist only on lattice sites
occupied at random by Mn atoms.
The approach mostly used\cite{ku} to calculate stiffness constant
within a random version of the Hubbard model was based on the
random phase approximation, where the electron-electron
approximation was taken into account in the Hartree-Fock
approximation and the disorder in the CPA. It is therefore very
probable that within this approach the formation of magnetic
clusters can be reproduced; the formation of the clusters is thus
strongly enviromental-dependent. However, the calculation of the
spatial GF Eq.( \ref{eq25a}), for the model,  Eq.( \ref{eq94}), is
rather a long and nontrivial task and we must   avoid considering
 this problem here. We hope, nevertheless, that the description
of the disorder effects, as given above, gives a good first
approximation as far as the the irreducible Green functions
method is concerned. A more detail consideration of the state of
itinerant carriers in DMS including a more sophisticated
treatment of disorder effects will be carried out separately.
\section{Conclusions}
In summary, we have presented an analytical approach for treating
the spin quasi-particle dynamics of the generalized spin-fermion
model, which provides a basis for description of the physical
properties of magnetic and diluted magnetic semiconductors. We
have investigated the influence of the correlation and exchange
effects for   interacting systems of itinerant carriers and
localized spins using the ideas of quantum field theory for
interacting electron and spin systems on a lattice. The workable
and self-consistent IGF approach to the decoupling problem for the
equation-of-motion method for double-time temperature Green
functions has been presented. The main achievement of this
formulation was the derivation of the Dyson equation for
double-time retarded Green functions instead of causal ones. That
formulation permits to unify   convenient analytical properties
of retarded and advanced GF and   the formal solution of the
Dyson equation  which, in spite of the required approximations for
the self-energy, provides the correct functional structure of
single-particle GF.  The main advantage of the mathematical
formalism is brought out by showing how elastic scattering
corrections (generalized mean fields) and inelastic scattering
effects (damping and finite lifetimes) could be self-consistently
incorporated in a general and compact manner. In this paper, we
have  confined   ourselves to the elastic scattering corrections
and have not considered the damping effects. This approach gives
a workable scheme for   definition of relevant generalized mean
fields written in terms of appropriate correlators. A comparative
study of real many-body dynamics of the generalized spin-fermion
model   is important  to characterize the true quasi-particle
excitations and the role of magnetic correlations. It was shown
that the magnetic dynamics of the generalized spin-fermion model
can be understood in terms of combined dynamics of itinerant
carriers, and of localized spins and magnetic correlations of
various nature. These applications illustrate some of   subtle
details of the IGF approach and exhibit their physical
significance in a representative  form.\\
As it is seen, this treatment has advantages in comparison with
the standard methods of decoupling of higher order GFs within the
equation-of-motion approach, namely, the following:

  At the mean-field level, the GF  one obtains, is richer
than that following from the standard procedures. The generalized
mean fields represent all elastic scattering renormalizations in
a compact form.\\
  The approximations ( the decoupling ) are introduced at
a later stage with respect to other methods,   i.e.,  only into
the rigorously obtained self-energy.\\
  The physical picture of   elastic and inelastic
scattering processes in the interacting many-particle systems is
clearly seen at every stage of calculations, which is not the
case with the standard methods of decoupling.\\
  The main advantage of the whole method is the
possibility of a {\it self-consistent} description of
quasi-particle spectra and their damping in a unified and coherent
fashion.\\
  Many   results of the previous works
are reproduced mathematically  more simply.\\
Thus, this new picture of an interacting spin-fermion system  on a
lattice is far richer and gives more possibilities for analysis
of phenomena which can actually take place. In this sense, the
approach we suggest produces more advanced physical picture of
the quasi-particle many-body dynamics. Our main results reveal
the fundamental importance of the adequate definition of
generalized mean fields at finite temperatures, which results in a
deeper insight into the nature of quasi-particle states of the
correlated lattice fermions and spins. The key to an
understanding of the situation in DMS lies in the right
description of the interplay of interactions and disorder effects
for coupled spin and charge subsystems. Consequently, it is
crucial that the correct functional structure of generalized mean
fields was calculated in a closed and compact form. The detailed
consideration of the state of itinerant charge carriers in DMS
along this line will be considered separately.
%

%
%
%
\end{document}